\documentclass[11pt, reqno]{amsart}
\usepackage{graphicx,amsmath,amssymb,epsfig}
\long\def\comment#1{}
\long\def\comment#1{}
\newtheorem{theorem}{Theorem}

\newtheorem{lemma}{Lemma}

\theoremstyle{definition}

\newcommand{\be}{\begin{eqnarray}}
\newcommand{\ee}{\end{eqnarray}}

\newcommand{\ba}{\begin{array}}
\newcommand{\ea}{\end{array}}
\newcommand{\bs}{\begin{align}\begin{split}\nonumber}
\newcommand{\bsnumber}{\begin{align}\begin{split}}
\newcommand{\es}{\end{split}\end{align}}

\renewcommand{\(}{\left(}
\renewcommand{\)}{\right)}
\renewcommand{\[}{\left[}
\renewcommand{\]}{\right]}
\renewcommand{\hat}{\widehat}

\newcommand{\Gn}{\mathbb{G}_n}
\newcommand{\Gnk}{\mathbb{G}_{n_k}}

\newcommand{\Pn}{\mathbb{P}_n}

\newcommand{\Ep}{{\mathrm{E}}}
\newcommand{\En}{{\mathbb{E}_n}}
\newcommand{\Ena}{{\mathbb{E}_{n_a}}}
\newcommand{\Enb}{{\mathbb{E}_{n_b}}}
\newcommand{\Enk}{{\mathbb{E}_{n_k}}}
\newcommand{\LASSO}{\sqrt{\textrm{LASSO}}}
\newcommand{\conflvl}{\gamma}

\def\RR{ {\Bbb{R}}}

\def\x{{x}}
\def\supp{{\rm support}}

\newcommand{\semin}[1]{\phi_{{\rm min}}(#1)}
\newcommand{\semax}[1]{\phi_{{\rm max}}(#1)}
\renewcommand{\hat}{\widehat}
\renewcommand{\leq}{\leqslant}
\renewcommand{\geq}{\geqslant}

\begin{document}

\title{Lasso Methods for Gaussian Instrumental Variables Models}
\author[Belloni \ Chernozhukov \ Hansen]{A. Belloni \and V. Chernozhukov \and C. Hansen}

\date{First version:  June 2009,  This \sc{version} of  \today.}

\maketitle

\begin{abstract} In this note, we propose the use of sparse methods (e.g. LASSO, Post-LASSO, $\LASSO$, and Post-$\LASSO$) to form first-stage predictions and estimate optimal instruments in linear instrumental variables (IV) models with many instruments in the canonical Gaussian case. The methods apply even when the number of instruments is much larger than the sample size. We derive asymptotic distributions for the resulting
IV estimators and provide conditions under which these sparsity-based IV estimators are asymptotically oracle-efficient.  In simulation experiments, a sparsity-based IV estimator with a data-driven penalty performs well compared to recently advocated many-instrument-robust procedures. We illustrate the procedure in an empirical example using the Angrist and Krueger (1991) schooling data.
 \\

\end{abstract}

\section{Introduction}
Instrumental variables (IV) methods are widely used in applied statistics, econometrics, and more generally for estimating treatment effects in situations where the treatment status is not randomly assigned; see, for example, \cite{acemoglu:colonial,AIR1996,AK2001,BaiNg2009,chen:sethi,Holland1986,NewhouseMcClellan,PermuttHebel,shimer:ur,SommerZeger}  among many others.  Identification of the causal effects of interest in this setting may be achieved through the use of observed instrumental variables that are relevant in determining the treatment status but are otherwise unrelated to the outcome of interest.  In some situations, many such instrumental variables are available, and the researcher is left with the question of which set of the instruments to use in constructing the IV estimator.  We consider one such approach to answering this question based on sparse-estimation methods in a simple Gaussian setting.

Throughout the paper we consider the Gaussian simultaneous equation model:\footnote{In a companion paper, \cite{BellChernHans:nonGauss} we consider the important generalization to heteroscedastic, non-Gaussian disturbances.  Focusing on the canonical Gaussian case allows for an elementary derivation of results, considerably sharper conditions, and much more refined penalty selection.
Therefore, the results for this canonical Gaussian case are of interest in their own right.}
\begin{eqnarray}
& &  y_{1i}  = y_{2i} \alpha_{1} + w_i'\alpha_2 + \epsilon_i, \label{Def: second stage} \\
& & y_{2i}  = D(x_i) + v_i,  \label{Def: first stage} \\
&  &  \(\begin{array}{cc} \epsilon_i \\ v_i\end{array}\)    \sim N\(0,\(\begin{array}{cc} \sigma^2_\epsilon & \sigma_{\epsilon v} \\ \sigma_{\epsilon v} & \sigma^2_{v}\end{array}\)\) \label{Def: errors}
\end{eqnarray}
where $y_{1i}$ is the response variable, $y_{2i}$ is the endogenous variable,  $w_i$ is a $k_w$-vector of control variables,
and $x_i = (z_i',w_i')'$ is a vector of instrumental variables (IV), and $(\epsilon_i, v_i)$ are disturbances that
are independent of $x_i$.
 The function $D(x_i) = \Ep[y_{2i}|x_i]$ is an unknown, potentially complicated function of the instruments.   Given a sample $(y_{1i}, y_{2i}, x_i), i =1,\ldots,n$, from the model above, the problem is to construct an IV estimator
for $\alpha_0 = (\alpha_1, \alpha_2')'$ that enjoys good finite sample properties and is asymptotically efficient.

We consider the case of fixed design, namely we treat the covariate values $x_1,\ldots,x_n$ as fixed. This includes random sampling as a special case; indeed, in this case $x_1,\ldots,x_n$ represent a realization of this sample on which we condition throughout. Note that for convenience, the notation has been collected in Appendix A.

First note that an asymptotically efficient, but infeasible, IV estimator
for this model takes the form
$$
\hat \alpha_I = \En[A_i d_i']^{-1} \En[A_i y_{1i}], \ \  A_i = ( D(x_i), w_i' )', \ \ d_i = (y_{2i}, w_i')'.
$$
Under suitable conditions,
$$
(\sigma^2_\epsilon Q^{-1}_n)^{-1/2}\sqrt{n}(\hat \alpha_I - \alpha_0) =_d  N(0,  I) + o_P(1)
$$
where $Q_n = \En [A_i A_i']$.

We would like
to construct an IV estimator that is as efficient as the infeasible optimal IV estimator $\hat \alpha_I$.
However, the optimal instrument $D(x_i)$ is an unknown function in practice and has to be estimated. Thus,
we investigate
estimating the optimal instruments $D(x_i)$ using sparse estimators arising from $\ell_1$-regularization procedures such as LASSO, post-LASSO, and others; see \cite{BickelRitovTsybakov2009,MY2007,CandesTao2007,BC-PostLASSO,BCW-SqLASSO}.  Such procedures are highly effective for estimating conditional expectations,
both computationally and theoretically,\footnote{Several $\ell_1$-regularized problems can be cast as convex programming problems and thus avoid the computational curse of dimensionality that would arise from a combinatorial search over models.} and, as we shall argue, are also effective for estimating optimal instruments.

In order to approximate the optimal instrument $D(x_i)$, we consider a large list of technical instruments,
\begin{equation}
f_i := (f_{i1},...,f_{ip})' := ( f_1(x_i),...,f_p(x_i))',
\end{equation}
where the number of instruments $p$ is possibly much larger than the sample size $n$.
High-dimensional instruments $f_i$ could arise easily because
\begin{itemize}
\item[(i)] the list of available instruments
is large, in which case, $f_i=x_i$,
\item[(ii)] or $f_i$ consists of a large number of series terms with respect to some elementary regressor vector $x_i$, e.g.,
B-splines, dummies, and/or polynomials, along with various interactions.
\end{itemize}
Without loss of generality we normalize the regressors so that $\En[f_{ij}^2]=1$ for $j=1,\ldots,p$.

The key condition that allows effective use of this large set of instruments is approximate sparsity which requires that most of the information in the optimal instrument can be captured by a relatively small number of technical instruments. Formally, approximate sparsity can be represented by the expansion of $D(x_i)$ as
\begin{eqnarray}\label{eq: sparse instrument}
&& D(x_i) = f_i' \beta_{0} + a(x_i),  \ \ \ \sqrt{\En [a(x_i)^2]} \leq c_{s} \lesssim \sigma_v\sqrt{s/n} , \ \ \ \|\beta_{0}\|_0  = s= o(n)
\end{eqnarray}
where the main part  $f_i' \beta_{0}$ of the optimal instrument uses only $s \ll n$ instruments, and the remainder
term $a(x_i)$ is approximation error that vanishes as the sample size increases.

The approximately sparse model (\ref{eq: sparse instrument}) substantially generalizes the classical parametric model of optimal instruments of \cite{amemiya:optimalIV} by letting
the identities of the relevant instruments
$$
T = \text{support}(\beta_{0})=\{ j \in \{1,\ldots,p\} \ : \ |\beta_{{0} j}| > 0\}
$$
be unknown and by allowing for approximation error in the parametric model for $D(x_i)$. This generalization is useful in practice since we do not know the identities of the relevant instruments in many examples. The model (\ref{eq: sparse instrument}) also generalizes the nonparametric model of optimal instruments of \cite{newey:optimaliv} by letting
the identities
of the most important series terms, $ T = \text{support}(\beta_{0})$, be unknown.  In this case, the number $s$ is
defined so that the approximation error is of the same order as the estimation error,
$\sqrt{s/n}$, of the oracle estimator. This rate generalizes the rate for the optimal
number $s$ of series terms in \cite{newey:optimaliv} by not relying on knowledge of what $s$ series terms to include.
Knowing
the identities of the most important series terms is unrealistic
in many examples in practice. Indeed, the most important series terms
need not be the first $s$ terms, and the optimal number
of series terms to consider is also unknown. Moreover, an optimal series approximation to the instrument
could come from the combination of completely different bases e.g
by using both polynomials and B-splines.

Based on the technical instruments $f_1,...,f_p$ and a sparse method such as LASSO or post-LASSO, we obtain estimates of $D(x_i)$ of the form
 \begin{equation}\label{Def: IV-LASSO1}
\widehat D(x_i) = f_i'\widehat \beta.
 \end{equation}
Sparse-methods take advantage of the approximate sparsity and ensure that many elements of $\widehat\beta$ are zero when $p$ is large. In other words, sparse-methods will select a small subset of the available instruments.  We then set
 \begin{equation}\label{Def: IV-LASSO2}
\widehat A_i =  ( \widehat D (x_i), w_i')'
 \end{equation}
to form the IV estimator
 \begin{equation}\label{Def: IV-LASSO3}
\widehat \alpha^* =  \Big(\En [\widehat A_i d_i']  \Big)^{-1}  \Big( \En [\widehat A_i y_{1i}]    \Big).
 \end{equation}

The main result of this note is to show that sparsity-based methods can produce estimates
of the optimal instruments $\widehat D_i$ based on a small, data-dependent set of instruments such that
 \begin{equation}
(\sigma^2_\epsilon Q^{-1}_n)^{-1/2} \sqrt{n}(\hat \alpha^* - \alpha_0) \to_d N(0, I)
 \end{equation}
under suitable regularity conditions.  That is, the IV estimator based on estimating the first-stage with appropriate sparse methods is asymptotically as efficient as the infeasible optimal IV estimator thus uses $D(x_i)$ and thus achieves the  semi-parametric efficiency bound.

Sufficient conditions for showing the IV estimator obtained using sparse-methods to estimate the optimal instruments is asymptotically efficient include a set of technical conditions and the following key growth condition:
$$ s^2 \log^2 p = o(n).$$
This rate condition requires the optimal
instruments to be sufficiently smooth  so that a small number of series terms can be used
to approximate them well.  This smoothness ensures that the impact of instrument estimation on the IV estimator is asymptotically negligible.

The rate condition above is substantive and can not be substantially weakened for the full-sample IV estimator considered above.  However,
we can replace this condition with the weaker condition that
$$ s \log p = o(n)$$
by employing a sample splitting method.  Specifically, we consider dividing the sample into (approximately) equal random parts $a$ and $b$, with sizes $n_a = \lceil n/2 \rceil$ and  $n_b = n - n_a$. We use superscripts $a$ and $b$ for variables in the first and second subsample respectively.  The index $i$ will enumerate observations in both samples, with ranges for the index given by $1 \leq i \leq n_a$ for sample $a$ and $1 \leq i \leq n_b$ for sample $b$. Let $\hat\sigma^{k}_j = \Enk[f_{ij}^{2}]^{1/2}$, $k=a,b$, $j=1,\ldots,p$, and $H_k = {\rm diag}(\hat\sigma_1^k,\ldots,\hat\sigma_p^k)$. Then we shall normalize the technical regressors in the subsamples, $f_{ij}^a = f_{ij}/\hat\sigma_j^a$, $f_{ij}^b = f_{ij}/\hat\sigma_j^b$, so that $\Ena[f_{ij}^{a2}]=1$ for and $\Enb[f_{ij}^{b2}]=1$ for $j=1,\ldots,p$.
 We can use each of the subsamples to fit the first stage via LASSO and variants, obtaining the first stage estimates $\hat \beta^k, k=a, b$. Then setting $\hat D_i^a = {f_i^a}'H_aH_b^{-1}\widehat \beta^b, 1 \leq i \leq n_a $,  $\hat D_i^b = {f_i^b}'H_bH_a^{-1}\widehat \beta^a, 1 \leq i \leq n_b$,  $\widehat A^k_i  = ( {\widehat{D}_i^{k}}, {w^k_i}')', k=a, b $,
we form the IV estimates in the two subsamples:
 \begin{equation}\label{Def:SplitIV}
\widehat \alpha_a = \Ena [\hat A_i^a  {d_i^a}' ]^{-1} \Ena [\hat A_i^a y^a_{1i}] \ \ \  \widehat \alpha_b = \Enb [\hat A_i^b  {d_i^b}' ]^{-1} \Enb [{d_i^b} y^b_{1i}].
 \end{equation}
Then we combine the estimate into one
 \begin{equation}\label{Def:SplitIVcombined}
\widehat \alpha_{ab} = ( n_a  \Ena [\hat A_i^a  \hat A_i^{a}{}']  +  n_b \Enb [\hat A_i^b  \hat A_i^{b}{}']   )^{-1} (n_a  \Ena [\hat A_i^a  \hat A_i^{a}{}'] \widehat \alpha_a +   n_b \Enb [\hat A_i^b  \hat A_i^{b}{}'] \widehat \alpha_b   );
 \end{equation}
where under i.i.d. sampling and random design we can also take
  \begin{equation}
\widehat \alpha_{ab} = \frac{1}{2} \hat \alpha_a + \frac{1}{2} \hat \alpha_b.
 \end{equation}
The second main result is to show that
 \begin{equation}
(\sigma^2_\epsilon Q^{-1}_n)^{-1/2} \sqrt{n}(\widehat \alpha_{ab} - \alpha_0) \to_d N(0, I)
 \end{equation}
under suitable regularity conditions.  That is, the IV estimator based on estimating the first-stage with appropriate sparse methods and sample splitting is asymptotically as efficient as the infeasible optimal IV estimator thus uses $D(x_i)$ and thus achieves the  semi-parametric efficiency bound.

\section{Properties of the IV Estimator with a Generic Sparsity-Based Estimator of Optimal Instruments}

In this section, we establish our main result. Under a high-level condition on the rates of convergence of a generic sparsity-based estimator of the optimal instrument, we show that the IV estimator based on these estimated instruments is asymptotically as efficient as the infeasible optimal estimator. Later we verify that many estimators that arise from sparse methods satisfy this rate condition.  In particular, we show they are satisfied for LASSO, Post-LASSO, $\LASSO$, and Post-$\LASSO$.

\begin{theorem}[Generic Result on Optimal IV Estimation]\label{lemma: lead lemma} In the linear IV model
of Section 1, assume that $\sigma_v$, $\sigma_\epsilon$ and the eigenvalues of $Q_n=\En[A_iA_i']$ are bounded away from zero and from above uniformly in $n$.
Let $\widehat D_{i}= f_i'\hat \beta$
be a generic sparsity-based estimator of optimal instruments $D_i = D(x_i)$ that obeys as $n$ grows
\begin{eqnarray}\label{RateConditionI}
&&  \|f_i'\widehat \beta -f_i'\beta_0\|_{2,n} +  c_s + \left \|\Gn ( f_{i} \epsilon_i  ) \right \|_{\infty}  \| \widehat \beta - \beta_{0}\|_{1} = o_P(1).\label{RateConditionI}
\end{eqnarray}
Then the IV estimator based on the equation (\ref{Def: IV-LASSO3}) is $\sqrt{n}$-consistent and is asymptotically oracle-efficient, namely as $n$ grows:
$$
(\sigma^2_\epsilon Q_n^{-1})^{-1/2} \sqrt{n}(\widehat \alpha^* - \alpha_0) =_d N(0, I) + o_P(1),
$$
and the result continues to hold with $Q_n$ replaced by $\widehat Q_n= \En [\widehat A_i \widehat A_i']$,
 and $\sigma^2_\epsilon$ by $\hat \sigma^2_\epsilon = \En[ (y_{1i} - \widehat A_i'\widehat \alpha^*)^2]$.
\end{theorem}

Theorem \ref{lemma: lead lemma} establishes the sufficiency of  (\ref{RateConditionI})  to derive the asymptotic oracle-efficiency of the proposed IV estimator. Under normality of the disturbances $\epsilon_i$, and standardized $f_{ij}$'s, we have
$$ \| \Gn( f_i \epsilon_i ) \|_\infty \lesssim_P \sigma_\epsilon\sqrt{ \log p }.$$
Thus, we shall have that (\ref{RateConditionI}) holds provided that $ s^2\log^2 p = o(n)$ by combining the relation above, standard approximation conditions (\ref{eq: sparse instrument}), and typical rates of convergence for sparse estimators (as shown in the next section).  The remaining conditions are quite standard and simply ensure that the optimal instruments would be well-behaved instrumental variables if they were known.

While the conditions of Theorem \ref{lemma: lead lemma} are quite general, we can weaken the sufficient rate condition by employing the split-sample IV estimator described in (\ref{Def:SplitIV}) and (\ref{Def:SplitIVcombined}).

\begin{theorem}[Generic Result on Optimal IV Estimation via Sample-Splitting]\label{lemma: lead lemma Split} In the linear IV model
of Section 1, assume that $\sigma_v$, $\sigma_\epsilon$ and the eigenvalues of $Q_n=\En[A_iA_i']$ are bounded away from zero and from above uniformly in $n$. Suppose that for  the generic split-sample estimates described in (\ref{Def:SplitIV})

\begin{eqnarray}\label{RateConditionSplitIV}
&&  \|\widehat D^k_i - D^k_i\|_{2,n_k} = o_P(1), \ \ \  k=a, b.
\end{eqnarray}
Then the split-sample IV estimator based on the equation (\ref{Def:SplitIVcombined}) is $\sqrt{n}$-consistent and is asymptotically oracle-efficient, namely as $n$ grows:
$$
(\sigma^2_\epsilon Q_n^{-1})^{-1/2} \sqrt{n}(\widehat \alpha_{ab} - \alpha_0) =_d N(0, I) + o_P(1),
$$
and the result continues to hold with $Q_n$ replaced by $\widehat Q_n= \En [\widehat A_i \widehat A_i']$
 and $\sigma^2_\epsilon$ by $\hat \sigma^2_\epsilon = \En[ (y_{1i} - \widehat A_i'\widehat \alpha_{ab})^2]$.
\end{theorem}

The conditions used in Theorem \ref{lemma: lead lemma Split} are quite similar to those in the conditions of Theorem \ref{lemma: lead lemma}.  The key difference is that the key condition (\ref{RateConditionSplitIV}) may obtain under the weaker rate condition that $s \log p = o(n)$.  Intuitively, weakening of the rate condition is due to the fact that using the first-stage coefficients estimated in one subsample to form estimates of optimal instruments in the other reduces the overfitting bias that drives the bias and inconsistency of two-stage least squares with many instruments.  Removing this bias allows one to efficiently estimate the second stage relationship while using more instruments than in a case where the overfitting bias is not controlled for.  It is important to note that the practical gains may be offset by the fact that the split-sample IV estimator avoids overfitting bias by fitting the first-stage on a much smaller set of observations than the full-sample procedure which generically produces a weaker first-stage relationship.  Thus, one is potentially trading off overfitting bias for weaker instruments and potential weak identification problems.  This tradeoff suggests the split-sample approach may perform relatively worse than the full-sample estimator in situations where instruments are not very strong in finite samples.

\section{Examples of Sparse Estimators of Optimal IV: LASSO and some of its modifications}

Given the sample $\{( y_{2i}, x_i), i=1,\ldots,n\}$ obeying the regression equations (\ref{Def: first stage}), we consider
estimators of the optimal instrument $D_i = D(x_i)$ that take
the form
\begin{equation}
\widehat D_i = \widehat D(\x_i) = f_i'\hat \beta,
 \end{equation}
where $\hat \beta$ is obtained by using a sparse method estimator with $y_{2i}$ as the dependent variable
and $f_i$ as regressors.

Recall that we consider the case of fixed design. Thus, we treat the covariate values $f_1,\ldots,f_n$ as fixed since the analysis is conditional on $x_1,\ldots,x_n$. Also, the dimension $p=p_n$ of each $f_i$  is allowed to grow as the sample size increases, where potentially $p>n$. In making asymptotic statements we also assume that $p\to \infty$ as $n\to \infty$.

Without loss of generality we normalize the regressors so that $\En[f_{ij}^2]=1$ for $j=1,\ldots,p$.

The classical AIC/BIC type estimators (\cite{Akaike1974,Schwarz1978}) are sparse as they solve the following optimization problem:
$$
\min_{\beta \in \Bbb{R}^p} \widehat Q (\beta) + \frac{\lambda}{n}  \|\beta\|_{0},
$$
where  $ \widehat Q (\beta) = \En[(y_{2i} - f_i'\beta)^2]$  and $\lambda$ is the penalty level. Unfortunately this problem is computationally prohibitive when $p$ is large, since the solution to the problem may require solving $\sum_{k \leqslant n} \binom{p}{k}$ least squares problems (thus, the complexity of this problem is NP-hard \cite{Natarajan1995,GeJiangYe2010}).

However, replacing the $\|\cdot\|_0$-regularization by a $\|\cdot\|_1$-regularization still yields sparse solutions and preserves the convexity of the criterion function. The latter substantially reduces the computational burden and makes these methods (provably) applicable to very high-dimensional problems. This method is called LASSO. In the following, we discuss LASSO and several variants in more detail.

\textbf{1. LASSO.} The LASSO estimator $\widehat \beta_L$ solves the following convex optimization problem:
        $$
\widehat \beta_L \in \arg\min_{\beta \in \Bbb{R}^p} \widehat Q (\beta) + \frac{\lambda}{n}  \|\beta\|_{1}.
$$
\begin{equation}\label{Def:LambdaLASSO} \mbox{with} \ \ \ \lambda = c \cdot  2\sigma_v \Lambda(1-\conflvl|F)\end{equation}
where $c>1$ (we recommend $c=1.1$) and $\Lambda(1-\conflvl|X)$ is the $(1-\conflvl)$-quantile of $$n\|\En[f_ig_i]\|_\infty$$ conditional on $F=[f_1,\ldots,f_n]'$, with $g_i \sim N(0,1)$ independent for $i=1,\ldots, n$. We note that $\Lambda(1-\conflvl|F) \leq \sqrt{n}\Phi^{-1}(1-\conflvl/2p) \leq \sqrt{2 n \log(p/\conflvl)}$. We set $\conflvl = 1/p$ which leads to $\conflvl=o(1)$ since $p\to \infty$ as $n\to\infty$.

\textbf{2. Post-LASSO.} The Post-LASSO estimator is simply ordinary least squares (OLS) applied to the data after removing the instruments/regressors that were not selected by LASSO. Set $$\widehat T_L = \supp( \hat \beta_{L} ) = \{ j \in \{1,\ldots,p\} \ : \ |\hat\beta_{L j }| > 0\}. \   $$
Then the post-LASSO estimator $\hat\beta_{PL}$ is \begin{equation}\label{Def:TwoStep} \widehat \beta_{PL} \in \arg\min_{\beta \in \mathbb{R}^p}   \hat Q(\beta) \ \ : \ \ \beta_j = 0 \ \ \mbox{if} \ j \not\in \widehat T_L.
\end{equation}

\textbf{3. Square-root LASSO.}  The $\sqrt{{\rm LASSO}}$ estimator is defined as the solution to the following optimization problem:
\begin{equation}\label{Def:LASSOmod}
\widehat \beta_{SQ} \in  \arg\min_{\beta \in \mathbb{R}^p} \sqrt{\hat Q(\beta)} +\frac{\lambda}{n}\|\beta\|_1.\end{equation}
\begin{equation}\label{our penalty}
 \mbox{ with } \ \ \ \lambda = c \cdot  \widetilde \Lambda(1-\conflvl|F)
 \end{equation}
 where $c>1$ (we recommend $c=1.1$) and
$\widetilde \Lambda(1-\conflvl|F)$ denotes the $(1-\conflvl)$-quantile of $$n\|\En[f_ig_i]\|_\infty/\sqrt{\En[g_i^2]}$$ conditional on $f_1,\ldots,f_p$, with $g_i \sim N(0,1)$ independent for $i=1,\ldots, n$.
We set $\conflvl = 1/p$ which leads to $\conflvl = o(1)$ as $n\to\infty$ since $p\to\infty$ as $n$ grows.

\textbf{4. Post-square-root LASSO.} The post-$\LASSO$ estimator is OLS applied to the data after removing the instruments/regressors that were not selected by $\LASSO$. Set $$\widehat T_{SQ} = \supp( \hat \beta_{SQ} ) = \{ j \in \{1,\ldots,p\} \ : \ |\hat\beta_{SQ j }| > 0\}, \   $$
and define the post-$\LASSO$ estimator $\hat\beta_{PSQ}$ as \begin{equation}\label{Def:TwoStepSqrt} \widehat \beta_{PSQ} \in \arg\min_{\beta \in \mathbb{R}^p}   \hat Q(\beta) \ \ : \ \beta_j = 0 \ \ \mbox{if} \ j \not\in \widehat T_{SQ}.
\end{equation}

The LASSO and $\LASSO$ estimators rely on $\ell_1$-norm regularization. By penalizing the $\ell_1$-norm of the coefficients, each estimator shrinks its estimated coefficients towards zero relative to the OLS estimator. Moreover, the kink at zero of the $\ell_1$-norm induces the estimators to have many zero components (in contrast with $\ell_2$-norm regularization).

Penalizing by $\|\beta\|_1$ yields sparse solutions but also introduces a bias towards zero on the components selected to be non-zero. In order to alleviate this effect, the post-LASSO and post-$\LASSO$ estimators are defined as ordinary least square regression applied to  the model selected by the LASSO or $\LASSO$.  It is clear that the post-LASSO and post-$\LASSO$ estimators remove the shrinkage bias of the associated estimator when it perfectly selects the model. Under mild regularity conditions, we also have that post-LASSO performs at least as well as LASSO when $\hat{m}$ additional variables are selected or LASSO misses some elements of $\beta_0$.
We refer to \cite{BC-PostLASSO} for a detailed analysis of post-LASSO.

The key quantity in the analysis of LASSO's properties is the score, the gradient of $\widehat Q$ evaluated at the true parameter ignoring approximation error:
$$ S = 2\En[ f_i v_i ].$$
The penalty level $\lambda$ should be chosen so it dominates the noise. Namely, for some $c>1$ one should set
\begin{equation}\label{Def:LambdaLASSOideal}
\lambda \geq c n \|S\|_\infty.
\end{equation}
Unfortunately, this is not feasible since $S$ is not observed. However, we can choose $\lambda$ based on the quantiles of $\|S\|_\infty$ conditional on $f_1,\ldots,f_n$. Note that the components of $S$ are normal (potentially correlated), so its distribution can be easily simulated. Using the choice of penalty level (\ref{Def:LambdaLASSO}) for LASSO, it follows that (\ref{Def:LambdaLASSOideal}) occurs with probability $1-\conflvl$.

The proposed penalty choice for the LASSO estimator depends on the standard deviation $\sigma_v$ of the disturbances. Typically, $\sigma_v$ is unknown and must be estimated. Relying on upper bounds on $\sigma_v$ can lead to an overly-large penalty and thus may result in potentially missing relevant components of $\beta_0$. The estimation of $\sigma_v$ could be done as proposed in \cite{BC-LectureNotes} under mild conditions. The square-root LASSO aims to circumvent this limitation.

As in the case of LASSO, the key quantity determining the choice of the penalty level for $\LASSO$ is its score  -- in this case the gradient of $\sqrt{\hat Q}$
evaluated at the true parameter value $\beta = \beta_0$ ignoring approximation error:
$$ \tilde S:= 
 \frac{\En[f_i v_i]}{\sqrt{\En[v^2_i]}}. $$
Because of the normalization by $\sqrt{\En[v_i^2]}$, the distribution of the score $\tilde S$ does not depend on the unknown standard deviation $\sigma_v$ or the unknown true parameter value $\beta_0$.
Therefore, the score is pivotal with respect to these parameters, conditional on $f_1,...,f_n$.
Thus,
setting  the penalty level as (\ref{our penalty}), with probability $1-\conflvl$, we have $$ \lambda \geq c n \|\tilde S\|_\infty.$$ We stress that the penalty level in (\ref{our penalty}) is independent of $\sigma_v$, in contrast to (\ref{Def:LambdaLASSO}). The properties of the $\LASSO$ have been studied in \cite{BCW-SqLASSO} where bounds similar to LASSO on the prediction norm and sparsity were established.

\section{Properties of IV Estimator with LASSO-based Estimators of Optimal IV}

In this section establish various rates of convergence of the sparse methods described in the previous section. In making asymptotic statements we also assume that $p\to \infty$ as $n\to \infty$.

\subsection{Regularity Conditions for Estimating Conditional Expectations}

The key technical condition used to establish the properties of the aforementioned sparsity-methods for estimating conditional expectations concerns the behavior of the empirical Gram matrix
 $M= \En[f_if_i']$. This matrix is necessarily singular when $p > n$, so in principle it is not well-behaved.
However, we only need good behavior of certain moduli of continuity of the Gram matrix. The first modulus of continuity is called the restricted eigenvalues and is needed for LASSO and $\LASSO$. The second modulus is called the sparse eigenvalue and is needed for Post-LASSO and Post-$\LASSO$.

In order to define the restricted eigenvalue, first define the restricted set:
$$\Delta_{C} = \{\delta \in \Bbb{R}^p: \|\delta_{T^c}\|_{1} \leq C  \| \delta_{T}\|_{1},  \delta \neq 0\},$$
where $T = \supp(\beta_0)$, then the restricted eigenvalues of a Gram matrix $M$ takes the form:
 \begin{eqnarray}\label{RE}
    \kappa^2_{C}: = \min_{\delta \in \Delta_{C}} s\frac{\delta' M \delta }{\|\delta_T\|^2_{1} } \text{ and }   \  \widetilde \kappa^2_{C} : = \min_{ \delta \in \Delta_{C} } \frac{\delta' M \delta }{\|\delta\|^2_2}.
\end{eqnarray}
These restricted eigenvalues can depend on $n$, but we suppress the dependence in our notation.

In making simplified asymptotic statements, we will invoke the following condition:

\textbf{Condition RE.} \textit{ For any $C>0$, there exists a finite constant $\kappa> 0$, which can depend on $C$, such that the restricted eigenvalues obey $\kappa_{C} \geq \kappa$  and $\widetilde \kappa_{C} \geq \kappa$ as $n \to \infty$.}

The restricted eigenvalue (\ref{RE}) is a variant of the  restricted eigenvalues introduced in Bickel, Ritov and Tsybakov \cite{BickelRitovTsybakov2009} to analyze the properties of LASSO in the classical Gaussian regression model.  Even though the minimal eigenvalue of the empirical Gram matrix $M$ is zero whenever $p > n$, \cite{BickelRitovTsybakov2009} show that its restricted eigenvalues can in fact be bounded away from zero. Many more
sufficient conditions are available from the literature; see \cite{BickelRitovTsybakov2009}.  Consequently, we take the restricted eigenvalues as primitive quantities and Condition RE as a primitive condition. Note also that the restricted eigenvalues are tightly tailored to the $\ell_1$-penalized estimation problem.

In order to define the sparse eigenvalues, let us define the $m$-sparse subset of a unit sphere as
$$
\Delta(m) = \{ \delta \in \Bbb{R}^p: \|\delta\|_0\leq m, \|\delta\|_2 = 1\},
$$
and also define the minimal and maximal $m$-sparse eigenvalue of the Gram matrix $M$ as
 \begin{equation}\label{Def:EigSparse}
\semin{m}  = \min_{\delta \in \Delta(m)}  \delta'M\delta \ \ \mbox{and} \ \
\semax{m}  = \max_{\delta \in \Delta(m)}  \delta'M\delta.
\end{equation}

To simplify asymptotic statements, we use the following condition:

\textbf{Condition SE.} \textit{ For any $C>0$, there exist constants
$0< \kappa' <  \kappa'' < \infty$  that do not depend on $n$ but can depend on $C$, such
that
$\kappa' \leq \semin{Cs}  \leq \semax{Cs}  \leq \kappa''$ as $n \to \infty$.
}

Condition SE requires that ``small" $m \times m$ submatrices of the large $p \times p$
empirical Gram matrix are well-behaved.
Moreover, Condition SE implies Condition RE by the argument given in \cite{BickelRitovTsybakov2009}.

It is well known that Conditions RE and SE are quite plausible for both many instrument and many series instrument settings. For instance, Conditions RE and SE hold for $M=\En[f_if_i']$ with probability approaching one as $n \to \infty$ if $f_{i}$ is a normalized form of $\tilde f_i$, namely   $f_{ij}= \tilde f_{ij}/\sqrt{\En[\tilde f_{ij}^2]}$, and
\begin{itemize}
\item $\tilde f_i$, $i = 1,\ldots,n$, are i.i.d. zero-mean Gaussian random vectors with population Gram matrix $\Ep[\tilde f_i \tilde f_i']$ has ones on the diagonal, its $s\log n$-sparse eigenvalues bounded from above and away from zero, and $s\log n = o(n/\log p)$;
\item $\tilde f_i$, $i=1,\ldots,n$, are i.i.d. bounded zero-mean random vectors with $\| \tilde f_i\|_\infty \leq K_n$ a.s. with population Gram  matrix $\Ep[\tilde f_i \tilde f_i']$ has ones on the diagonal, its $s\log n$-sparse eigenvalues bounded from above and away from zero, $\sqrt{n}/K_n \to \infty$, and $ s\log n = o((1/K_n)\sqrt{n/\log p})$.
\end{itemize}

Recall that a standard assumption in econometric research is to assume that the population Gram matrix $\Ep[f_i f_i']$ has eigenvalues bounded from above and below, see e.g. \cite{newey:series}.
The conditions above allow for this and much more general behavior, requiring only that the sparse eigenvalues of the population Gram matrix $\Ep[f_i f_i']$ are bounded from below and from above. The latter is important for allowing functions $f_i$ to be formed as a combination of elements from different bases, e.g. a combination of B-splines with polynomials.  The lemmas above further show that under some restrictions on the growth of $s$ in relation to the sample size $n$, the good behavior of the population sparse eigenvalues translates into a good behavior of empirical sparse eigenvalues, which ensures that Conditions RE and SE are satisfied
in large samples.

\subsection{Results on Sparse Estimators under Gaussian Errors}

Next we gather rate of convergence results for the different sparse estimators discussed in Section 3. We begin with the rates for LASSO and Post-LASSO.

\begin{lemma}[Rates for LASSO and Post-LASSO]\label{RatesLASSO} Suppose we have the sample of size $n$ from the model  $y_{2i}  = D(x_i) + v_i, i =1,..,n$ where $x_i, i=1,...,n$ are fixed, and $v_i,i=1,...,n$ are i.i.d Gaussian with variance $\sigma^2_v$.  Suppose that the approximate sparsity condition (\ref{eq: sparse instrument}) holds for the function $D(x_i)$ with respect to $f_i$, and that Conditions RE and SE hold for $M = \En[f_i f_i']$. Suppose the penalty level for LASSO is specified as in (\ref{Def:LambdaLASSO}) with $\conflvl = o(1)$ as $n$ grows. Then, as $n$ grows, for $\widehat \beta$ defined as either the LASSO or Post-LASSO estimator
and the associated fit $\hat D_{i} = f_i' \hat \beta$
\begin{eqnarray*}
& \displaystyle \|\hat D_{i}  - D_{i}\|_{2,n}  & \lesssim_P  \sigma_v\sqrt{\frac{s \log (p/\conflvl)}{ n }}, \\
& \displaystyle   \|\hat{\beta} - \beta_0 \|_2  & \lesssim_P \sigma_v\sqrt{\frac{s \log (p/\conflvl)}{ n }}, \\
& \displaystyle     \ \ \|\hat{\beta} - \beta_{0} \|_1  & \lesssim_P \sigma_v\sqrt{\frac{s^2 \log (p/\conflvl)}{ n }}.
  \end{eqnarray*}
 \end{lemma}

The following lemma derives the properties for $\LASSO$ and Post-$\LASSO$.

\begin{lemma}[Rates for $\LASSO$ and Post-$\LASSO$]\label{RatesSQLASSO}
Suppose we have the sample of size $n$ from the model  $y_{2i}  = D(x_i) + v_i, i =1,..,n$ where $x_i, i=1,...,n$ are fixed, and $v_i,i=1,...,n$ are i.i.d Gaussian with variance $\sigma^2_v$.  Suppose that the approximate sparsity condition (\ref{eq: sparse instrument}) holds for the function $D(x_i)$ with respect to $f_i$, and that Conditions RE and SE hold for $M = \En[f_i f_i']$.  Suppose the penalty level for $\LASSO$ is specified as in (\ref{our penalty}) with $\conflvl = o(1)$ as $n$ grows. Then, as $n$ grows, provided $s \log (p/\conflvl) = o(n)$,  $\widehat \beta$ defined as either the $\LASSO$ or Post-$\LASSO$ estimator
and the associated fit $\hat D_{i} = f_i' \hat \beta$ satisfy
\begin{eqnarray*}
&\displaystyle  \|\hat D_{i}  - D_{i}\|_{2,n}  & \lesssim_P  \sigma_v\sqrt{\frac{s \log (p/\conflvl)}{ n }}, \\
& \displaystyle   \|\hat{\beta} - {\beta}_0 \|_2  & \lesssim_P \sigma_v\sqrt{\frac{s \log (p/\conflvl)}{ n }}, \\
& \displaystyle   \|\hat{\beta} - {\beta}_0 \|_1  & \lesssim_P \sigma_v\sqrt{\frac{s^2 \log (p/\conflvl)}{ n }}.
  \end{eqnarray*}
 \end{lemma}

Although all these estimators enjoy similar rates, their practical performance in finite sample can be relatively different. As mentioned before, Post-LASSO aims to reduce the regularization bias introduced by LASSO. This is typically desirable if LASSO generated a sufficiently sparse estimator so that Post-LASSO does not overfit. However, LASSO (and therefore Post-LASSO) relies on the knowledge or pre-estimation of the standard deviation $\sigma_v$ of the disturbances $v_i$. $\LASSO$ circumvent that at the cost of a mild side condition having to hold (typically $s\log p = o(n)$). Finally, Post-$\LASSO$ aims to remove the shrinkage bias inherent in $\LASSO$.

Based on these lemmas we can achieve the results in Theorem \ref{lemma: lead lemma} based on primitive assumptions for either of these four estimators.

\begin{theorem}[Asymptotic Normality for IV Based on LASSO, Post-LASSO, $\LASSO$ and Post-$\LASSO$]
In the linear IV model
of Section 1, assume that $\sigma_v$, $\sigma_\epsilon$ and the eigenvalues of $Q_n=\En[A_iA_i']$ are bounded away from zero and from above uniformly in $n$. Suppose that the optimal instrument is approximately sparse, namely  (\ref{eq: sparse instrument}) holds,
  conditions RE, SE hold for $M=\En[f_i f_i']$, $\conflvl = 1/p = o(1)$, and $s^2\log^2 p = o(n)$ hold, and let $\widehat D_{i}= f_i'\hat \beta$
where $\hat \beta$ is the LASSO, Post-LASSO, $\LASSO$ or  Post-$\LASSO$ estimator.
Then the IV estimator based on the equation (\ref{Def: IV-LASSO3}) is $\sqrt{n}$-consistent and is asymptotically oracle-efficient, namely as $n$ grows:
$$
(\sigma^2_\epsilon Q_n^{-1})^{-1/2} \sqrt{n}(\widehat \alpha^* - \alpha_0) =_d N(0, I) + o_P(1),
$$
and the result continues to hold with $Q_n$ replaced by $\widehat Q_n= \En [\widehat A_i \widehat A_i']$,
 and $\sigma^2_\epsilon$ by $\hat \sigma^2_\epsilon = \En[ (y_{1i} - \widehat A_i'\widehat \alpha^*)^2]$.
\end{theorem}

In the analysis of split-sample IV recall that we re-normalized the technical regressors in the subsamples so that $\Ena[f_{ij}^{a2}]=1$ for $j=1,\ldots,p$ and $\Enb[f_{ij}^{b2}]=1$ for $j=1,\ldots,p$, and the LASSO estimators are applied to such samples. Letting $H_k = {\rm diag}(\hat\sigma_1^k,\ldots,\hat\sigma_p^k)$, we have from condition (\ref{eq: sparse instrument}) that
 \begin{equation}\label{eq: sparse instrument2}
 D(x^k_i) = f_i{}' \beta_{0} + a(x^k_i) = f^k_i{}' H_k\beta_{0} + a(x^k_i), \ \ k =a, b,
 \end{equation}
so that an approximate sparse model for each subsample follows from the approximate sparse model $\beta_0$ for the full sample times the appropriate diagonal matrix containing the normalizations needed to enforce the normalizations above.

\begin{theorem}[Asymptotic Normality for Split-Sample IV Based on LASSO, Post-LASSO, $\LASSO$ and Post-$\LASSO$]
In the linear IV model
of Section 1, assume that $\sigma_v$, $\sigma_\epsilon$ and the eigenvalues of $Q_n=\En[A_iA_i']$ are bounded away from zero and from above uniformly in $n$. Suppose that the optimal instrument is approximately sparse, namely  (\ref{eq: sparse instrument}) holds,
  conditions RE, SE hold for $M^k=\Enk[f^k_i f^k_i{}']$ for $k=a,b$, $\conflvl = 1/p = o(1)$, and $s\log p = o(n)$ hold, and let $\widehat D^k_{i}= f^k_i{}'H_kH_{k^c}^{-1}\hat \beta^{k^c}$
where $\hat \beta^{k^c}$ is the LASSO, Post-LASSO, $\LASSO$ or  Post-$\LASSO$ estimators applied to the subsample $ \{(y^{k^c}_{2i},f^{k^c}_i) : 1\leq i \leq n_{k^c}\}$ for  $k =a,b$, and $k^c=\{a,b\}\setminus k$.  Then the split-sample IV estimator based on the equation (\ref{Def:SplitIVcombined}) is $\sqrt{n}$-consistent and is asymptotically oracle-efficient, namely as $n$ grows:
$$
(\sigma^2_\epsilon Q_n^{-1})^{-1/2} \sqrt{n}(\widehat \alpha_{ab} - \alpha_0) =_d N(0, I) + o_P(1),
$$
and the result continues to hold with $Q_n$ replaced by $\widehat Q_n= \En [\widehat A_i \widehat A_i']$
 and $\sigma^2_\epsilon$ by $\hat \sigma^2_\epsilon = \En[ (y_{1i} - \widehat A_i'\widehat \alpha_{ab})^2]$.
\end{theorem}

Theorems 3 and 4 verify that the conditions required in the generic results given in Theorems 1 and 2 are satisfied when LASSO, $\LASSO$, post-LASSO, or post-$\LASSO$ are used to estimate the optimal instruments.  The conditions in the two theorems are quite similar.  As mentioned above, the condition on sparsity embodied by the rate condition $s\log(p) = o(n)$ in Theorem 4 is weaker than the analogous condition in Theorem 3.  Both results also impose restrictions on the empirical design matrices, $M$ in Theorem 3 and $M_k$ for $k = a,b$ in Theorem 4.  These conditions are similar to, but weaker than, the usual full-rank condition for estimating linear models via ordinary least squares.  Both theorems also implicitly assume that identification is strong, i.e. that $D(x_i)$ is bounded away from the zero-function.

\section{Simulation Experiment}
The theoretical results presented in the previous sections suggest that using LASSO to aid in fitting the first-stage regression should result in IV estimators with good estimation and inference properties.  In this section, we provide simulation evidence on estimation and inference properties of IV estimators using LASSO and $\LASSO$ to select instrumental variables for a second-stage estimator.  All results reported in this section are for post-LASSO and post-$\LASSO$ but we refer to LASSO or $\LASSO$ to simplify the presentation.

Our simulations are based on a simple instrumental variables model of the form
\begin{align*}
y_i &= \alpha d_i + e_i \\
d_i &= z_i'\Pi + v_i \\
(e_i,v_i) &\sim N\(0,\(\begin{array}{cc} \sigma^2_e & \sigma_{ev} \\ \sigma_{ev} & \sigma^2_{v}\end{array}\)\) \ iid
\end{align*}
where $\alpha=1$ is the parameter of interest, and $z_i = (z_{i1},z_{i2},...,z_{i100})' \sim N(0,\Sigma_Z)$ is a 100 x 1 vector with $E[z_{ih}^2] = \sigma^2_z$ and $Corr(z_{ih},z_{ij}) = .5^{|j-h|}$.  In all simulations, we set $\sigma^2_e = 1$ and $\sigma^2_z = 1$.

For the other parameters, we use a variety of different parameter settings.  We provide simulation results for sample sizes, $n$, of 101 and 500.  We consider two different values for $Corr(e,v)$: .3 and .6.  We also consider three values of $\sigma^2_v$ which are chosen to benchmark three different strengths of instruments.  The three values of $\sigma^2_v$ are found as $\sigma^2_v = \frac{n \Pi'\Sigma_Z\Pi}{F^*\Pi'\Pi}$ for three different values of $F^*$: 10, 40, and 160.  Finally, we use two different settings for the first stage coefficients, $\Pi$.  The first sets the first five elements of $\Pi$ equal to one and the remaining elements equal to zero.  We refer to this design as the ``cut-off'' design.  The second model sets the coefficient on $z_{ih} = .7^{h-1}$ for $h=1,...,100$.  We refer to this design as the ``exponential'' design.  In the cut-off case, the first-stage has an exact sparse representation, while in the exponential design, the model is not literally sparse although the majority of explanatory power is contained in the first few instruments.

For each setting of the simulation parameter values, we report results from several estimation procedures.  A simple possibility when presented with $p < n$ instrumental variables is to just estimate the model using 2SLS and all of the available instruments.  It is well-known that this will result in poor-finite sample properties unless there are many more observations than instruments; see, for example, \cite{bekker}.  Fuller's estimator \cite{fuller} (FULL)\footnote{The Fuller estimator requires a user-specified parameter.  We set this parameter equal to one which produces a higher-order unbiased estimator.  See \cite{hhk:weakmse} for additional discussion.} is robust to many instruments as long as the presence of many instruments is accounted for when constructing standard errors for the estimators and $p < n$; see \cite{bekker} and \cite{hhn:weakiv} for example.  We report results for these estimators in rows labeled 2SLS(100) and FULL(100) respectively.  In addition, we report Fuller and IV estimates based on the set of instruments selected by LASSO or $\LASSO$ with two different penalty selection methods.  IV-LASSO, FULL-LASSO, IV-SQLASSO, and FULL-SQLASSO are respectively 2SLS and Fuller using instruments selected by LASSO and 2SLS and Fuller using instruments selected by $\LASSO$ using the simple plug-in penalties given in Section 3.  IV-LASSO-CV, FULL-LASSO-CV, IV-SQLASSO-CV, and FULL-SQLASSO-CV are respectively 2SLS and Fuller using instruments selected by LASSO and 2SLS and Fuller using instruments selected by $\LASSO$ using the 10-fold cross-validation to choose the penalty level.  For each estimator, we report root-mean-squared-error (RMSE), median bias (Med. Bias), mean absolute deviation (MAD), and rejection frequencies for 5\% level tests (rp(.05)).  For computing rejection frequencies, we estimate conventional 2SLS standard errors for all 2SLS estimators, and the many instrument robust standard errors of \cite{hhn:weakiv} for the Fuller estimators.

Simulation results are presented in Tables 1-4.  Tables 1-2 give results for the cut-off design with $n = 101$ and $n = 500$ respectively; and Tables 3-4 give results for the exponential design with $n = 101$ and $n = 500$ respectively. As expected, 2SLS(100) does extremely poorly along all dimensions.  FULL(100) also performs worse than the LASSO- and $\LASSO$-based estimators in terms of estimator risk (RMSE and MAD) in all cases.  With $n = 500$, FULL(100) is on par with the LASSO- and $\LASSO$-based estimators in terms of rejection frequencies (rp(.05)) but tends to perform much worse than these with $n = 101$.

All of the LASSO- and $\LASSO$-based procedures perform similarly in the two examples with $n = 500$.  Outside of outperforming the two procedures that use all of the instruments, there is little that systematically differentiates the various estimators looking at RMSE and MAD. There appears to be a tendency for the estimates with variable selection done with the simple plug-in penalty to have slightly smaller RMSE and MAD than the estimates based on using cross-validation to choose the penalty, though the pattern is not striking.  Looking at median bias, the Fuller estimator has uniformly smaller bias than the associated 2SLS estimator that uses the same instruments as predicted by the theory for the Fuller estimator.  That this does not equate to uniformly smaller MAD or RMSE is of course due to the fact that Fuller estimator is slightly more variable than 2SLS.  Finally, all estimators do fairly well in terms of 95\% coverage probabilities, though once again the Fuller-based tests have uniformly smaller size-distortions than the associated 2SLS tests using the same instruments.  Tests that use instruments selected by cross-validation also do worse in terms of coverage probabilities than the tests that use the simple plug-in rule.  This difference is especially pronounced for small values of F* and goes away as F* increases.

The major qualitative difference in the results for the LASSO- and $\LASSO$-based procedures with $n = 101$ as compared with $n = 500$ are in the numbers of cases in which the variable selection methods choose no instruments.  With $n = 101$, we see that $\LASSO$ tends to be more conservative in instrument selection than LASSO and that, unsurprisingly, CV tends to select more variables than the more conservative plug-in rule.   For example, with F* = 10 and Corr(e,v) = .3 in the exponential design, LASSO and $\LASSO$ with the plug-in penalty select no instruments in 122 and 195 cases while LASSO and $\LASSO$ using 10-fold cross-validation select no instruments in only 27 and 30 cases.  Outside of this, the same basic patterns for RMSE, MAD, median bias, and rejection probabilities discussed in the $n = 500$ case continue to hold.

Overall, the simulation results are favorable to the LASSO- and $\LASSO$-based IV methods.  The LASSO- and $\LASSO$-based estimators dominate the other estimators considered based on RMSE or MAD and have relatively small finite sample biases.  The LASSO- and $\LASSO$-based procedures also do a good job in producing tests with size close to the nominal level.  There is some evidence that the Fuller estimator using instruments selected by LASSO may do better than the more conventional 2SLS estimator in terms of testing performance.  In the designs considered, it also seems that the simple plug-in rule may produce estimates that behave slightly better than those obtained by using cross-validation to choose the LASSO and $\LASSO$ penalty levels.  It may be interesting to explore these issues in more depth in future research.

\section{Instrument Selection in Angrist and Krueger Data}

Next we apply post-LASSO in the Angrist and Krueger \cite{AK1991} model $$
\begin{array}{lll}
 y_{i1}  & = \theta_1 y_{i2} +
w_i'\gamma + \epsilon_i, &  \Ep[\epsilon_i|w_i, x_i] = 0,\\
 y_{i2}  & = z_i'\beta + w_i'\delta +
v_i,  &  \Ep[v_i|w_i, x_i] = 0,
\end{array}
 $$
where  $y_{i1}$ is the log(wage) of individual $i$,  $y_{i2}$ denotes education, $w_i$ denotes a vector of control variables, and $x_i$ denotes a vector of instrumental variables that affect education but do not directly affect the wage. The data were drawn from the 1980 U.S. Census and consist of 329,509 men born between 1930 and 1939.  In this example, $w_i$ is a set of 510 variables: a constant, 9 year-of-birth dummies, 50 state-of-birth dummies, and 450 state-of-birth $\times$ year-of-birth interactions.  As instruments, we use three quarter-of-birth dummies and interactions of these quarter-of-birth dummies with the set of state-of-birth and year-of-birth controls in $w_i$ giving a total of 1530 potential instruments.  \cite{AK1991} discusses the endogeneity of schooling in the wage equation and provides an argument for the validity of $z_i$ as instruments based on compulsory schooling laws and the shape of the life-cycle earnings profile.  We refer the interested reader to \cite{AK1991} for further details.  The coefficient of interest is $\theta_1$, which summarizes the causal impact of education on earnings.

There are two basic options that have been used in the literature: one uses just the three basic quarter-of-birth dummies and the other uses 180 instruments corresponding to the three quarter-of-birth dummies and their interactions with the 9 main effects for year-of-birth and 50 main effects for state-of-birth.  It is commonly-held that using the set of 180 instruments results in 2SLS estimates of $\theta_1$ that have a substantial bias, while using just the three quarter-of-birth dummies results in an estimator with smaller bias but a larger variance; see, e.g., \cite{hhn:weakiv}.  Another approach uses the 180 instruments and the Fuller estimator \cite{fuller}  (FULL) with an adjustment for the use of many instruments.  Of course, the sparse methods for the first-stage estimation explored in this paper offer another option that could be used in place of any of the aforementioned approaches.

Table 5 presents estimates of the returns to schooling coefficient using 2SLS and FULL\footnote{We set the user-defined choice parameter in the Fuller estimator equal to one which results in a higher-order unbiased estimator.} and different sets of instruments.  Given knowledge of the construction of the instruments, the first three rows of the table correspond to the natural groupings of the instruments into the three main quarter of birth effects, the three quarter-of-birth dummies and their interactions with the 9 main effects for year-of-birth and 50 main effects for state-of-birth, and the full set of 1530 potential instruments.  The remaining two rows give results based on using LASSO to select instruments with penalty level given by the simple plug-in rule in Section 3 or by 10-fold cross-validation.\footnote{Due to the similarity of the performance of LASSO and $\LASSO$ in the simulation, we focus only on LASSO results in this example.}  Using the plug-in rule, LASSO selects only the dummy for being born in the fourth quarter, and with the cross-validated penalty level, LASSO selects 12 instruments which include the dummy for being born in the third quarter, the dummy for being born in the fourth quarter, and 10 interaction terms.  The reported estimates are obtained using post-LASSO.

The results in Table 5 are interesting and quite favorable to the idea of using LASSO to do variable selection for instrumental variables.  It is first worth noting that with 180 or 1530 instruments, there are modest differences between the 2SLS and FULL point estimates that theory as well as evidence in \cite{hhn:weakiv} suggests is likely due to bias induced by overfitting the 2SLS first-stage which may be large relative to precision.  In the remaining cases, the 2SLS and FULL estimates are all very close to each other suggesting that this bias is likely not much of a concern.  This similarity between the two estimates is reassuring for the LASSO-based estimates as it suggests that LASSO is working as it should in avoiding overfitting of the first-stage and thus keeping bias of the second-stage estimator relatively small.

For comparing standard errors, it is useful to remember that one can regard LASSO as a way to select variables in a situation in which there is no \textit{a priori} information about which of the set of variables is important; i.e. LASSO does not use the knowledge that the three quarter of birth dummies are the ``main'' instruments and so is selecting among 1530 \textit{a priori} ``equal'' instruments.  Given this, it is again reassuring that LASSO with the more conservative plug-in penalty selects the dummy for birth in the fourth quarter which is the variable that most cleanly satisfies Angrist and Krueger's \cite{AK1991} argument for the validity of the instrument set.  With this instrument, we estimate the returns-to-schooling to be .0862 with an estimated standard error of .0254.  The best comparison is FULL with 1530 instruments which also does not use any \textit{a priori} information about the relevance of the instruments and estimates the returns-to-schooling as .1019 with a much larger standard error of .0422.  In the same information paradigm, one can be less conservative than the plug-in penalty by using cross-validation to choose the penalty level.  In this case, only 12 instruments are chosen producing a Fuller point estimate (standard error) of .0997 (.0139) or 2SLS point estimate (standard error) of .0982 (.0137).  These standard errors are smaller than even the standard errors obtained using information about the likely ordering of the instruments given by using 3 or 180 instruments where FULL has standard errors of .0200 and .0143 respectively.  That is, LASSO finds just 12 instruments that contain nearly all information in the first stage and, by keeping the number of instruments small, produces a 2SLS estimate that likely has relatively small bias.\footnote{Note that it is simple to modify LASSO to use \textit{a priori} information about the relevance of instruments by changing the weighting on different coefficients in the penalty function.  For example, if one uses the plug-in penalty and simultaneously decreases the penalty loading on the three main quarter of birth instrument to reflect beliefs that these are the most relevant instruments, one chooses only the three quarter of birth instruments.}  Overall, these results demonstrate that LASSO instrument selection is both feasible and produces sensible and what appear to be relatively high-quality estimates in this application.

\section*{Acknowledgement} We would like to thank Josh Angrist for helpful suggestions on the empirical example that led to the large number of instruments.

\appendix

\section{Notation.}   We allow
for the models to change with the sample size, i.e. we allow for array asymptotics.  Thus, all parameters are
implicitly indexed by the sample size $n$, but we omit the index to simplify
notation.  We use array asymptotics to better capture some finite-sample phenomena.
We also use the following empirical process notation, $$\En[f] = \En[f(z_i)] = \sum_{i=1}^n f(z_i)/n,$$ and $$\Gn(f) = \sum_{i=1}^n ( f(z_i)
- \Ep[f(z_i)] )/\sqrt{n}.$$
The ${l}_2$-norm is denoted by
$\|\cdot\|$, and the ${l}_0$-norm, $\|\cdot\|_0$, denotes the number of non-zero components of a vector.  The empirical $L^2(\Pn)$ norm
of a random variable $W_i$ is defined as
$$
\|W_i\|_{2,n} := \sqrt{\En[ W_i^2]}.
$$
Given a vector $\delta \in \RR^p$, and a set of
indices $T \subset \{1,\ldots,p\}$, we denote by $\delta_T$ the vector in which $\delta_{Tj} = \delta_j$ if $j\in T$, $\delta_{Tj}=0$ if $j \notin T$. We use the notation $(a)_+ = \max\{a,0\}$, $a \vee b = \max\{ a, b\}$ and $a \wedge b = \min\{ a , b \}$. We also use the notation $a \lesssim b$ to denote $a \leqslant c b$ for some constant $c>0$ that does not depend on $n$; and $a\lesssim_P b$ to denote $a=O_P(b)$. For an event $E$, we say that $E$ wp $\to$ 1 when $E$ occurs with probability approaching one as $n$ grows. We say $X_n =_d Y_n + o_P(1)$ to mean that $X_n$ has the same distribution as $Y_n$ up to a term $o_P(1)$ that vanishes in probability. Such statements are needed to accommodate asymptotics for models that change with $n$.  When $Y_n$ is a fixed random vector, that does not change with $n$, i.e. $Y_n = Y$, this notation is equivalent to $X_n \to_d Y$.

\section{Proof of Theorem 1}

Step 0. Recall that $A_i = (D_i,w_i')'$ and $d_i = (y_{2i},w_i')'$ for $i=1,\ldots,n$. The condition that $\En[A_iA_i'] = Q_n$ has eigenvalues bounded from above uniformly in $n$ implies that
 $$\En[D_i^2]+\En[\|w_i\|^2] = \En[\|A_i\|^2] = {\rm trace}(Q_n)\lesssim (1+k_w)$$
is bounded from above uniformly in $n$.

Also, we have $\En[A_i\epsilon_i] \sim N(0,\sigma_\epsilon^2Q_n/n)$ and $\En[A_iv_i] \sim N(0,\sigma_v^2Q_n/n)$ so that  $$\|\En[d_i\epsilon_i]\| \leq |\En[v_i\epsilon_i]| + \|\En[A_i\epsilon_i]\| \lesssim_P \sigma_{\epsilon v} + \sqrt{(1+k_w)/n}$$
$$ \|\En[A_iv_i]\|^2 = |\En[D_iv_i]|^2 + \|\En[w_iv_i]\|^2 \lesssim_P \sigma_v^2 (1+k_w)/n$$
$$ \|d_i\|_{2,n} \leq \|v_i\|_{2,n} + \|A_i\|_{2,n} \lesssim_P \sigma_v + \sqrt{1+k_w} $$
where $\sigma_v$ and $k_w$ are bounded from above uniformly in $n$.

Step 1. We have that by $\Ep[\epsilon_i|A_i]=0$
\begin{eqnarray*}
\sqrt{n}(\widehat \alpha^* - \alpha_0) &= & \En [\widehat A_i d_i']^{-1} \sqrt{n} \En [\widehat A_i \epsilon_i]  \\
& = & \{\En [\widehat A_i d_i']\}^{-1} \Gn [\widehat A_i \epsilon_i]  \\
& = & \{\En [A_i d_i'] + o_P(1)\}^{-1} \left( \Gn [A_i \epsilon_i] + o_P(1) \right)
 \end{eqnarray*}
where by Steps 2 and 3 below:
\begin{eqnarray}
 \En [\widehat A_id_i'] = \En[A_id_i'] + o_P(1) \label{eq: to show 1} \\
 \Gn [\widehat A_i \epsilon_i] = \Gn [A_i \epsilon_i] + o_P(1)\label{eq: to show 2}.
 \end{eqnarray}
Thus, since $\En[D_i(y_{2i}-D_i)] = o_P(1)$ and $\En[w_i(y_{2,i}-D_i)]=o_P(1)$ by Step 0, note that $\En[A_id_i'] = \En[A_i A_i'] + o_P(1) = Q_n + o_P(1)$. Moreover, by the assumption on $\sigma_\epsilon$ and $Q_n$, $Var(\Gn [ A_i \epsilon_i]) = \sigma^2_\epsilon Q_n$ has eigenvalues bounded away from zero and bounded from above, uniformly in $n$. Therefore,
$$
\sqrt{n}(\widehat \alpha^* - \alpha_0) = Q^{-1}_n  \Gn [A_i \epsilon_i] + o_P(1),
$$
and  $Q^{-1}_n  \Gn [A_i \epsilon_i]$ is a vector distributed as normal with mean zero and covariance
$\sigma^2_\epsilon Q^{-1}_n$.

Step 2.   To show (\ref{eq: to show 1}), note that $\widehat A_i - A_i = (\widehat D_i - D_i, 0')'$. Thus,
 \begin{eqnarray*}
\| \En[(\widehat A_i - A_i)d_i']\| = \| \En[(\widehat D_i - D_i)d_i ']\| &\leq&  \En[\|\widehat D_i-D_i\| \|d_i\|] \\
& \leq & \sqrt{\En\[ \|\widehat D_{i}-D_{i}\|^2\] \ \En[\|d_i\|^2]}\\
& = &  \|\widehat D_{i} -D_{i}\|_{2,n} \cdot \|d_i\|_{2,n} \\
& \lesssim_P &  \|\widehat D_{i} -D_{i}\|_{2,n} = o_P(1)
 \end{eqnarray*}
since  $ \|d_i\|_{2,n} \lesssim_P 1$ by Step 0, and the rate assumption (\ref{RateConditionI}).

Step 3.  To show (\ref{eq: to show 2}), note that
 \begin{eqnarray*}
  \|\Gn[(\widehat A_{i} - A_{i}) \epsilon_i]\|   & = & |\Gn[(\widehat D_{i} - D_{i}) \epsilon_i]|  \\
   & = & | \Gn \{f_i'(\widehat \beta - \beta_{0})  \epsilon_i \}  + \Gn \{ a_{i} \epsilon_i \} |  \\
 & = &   \left| \sum_{j=1}^p \Gn \{ f_{ij} \epsilon_i  \} '  (\widehat \beta_{j} - \beta_{0j})   + \Gn \{ a_{i} \epsilon_i \} \right|  \\
 & \leq &
\left\|\Gn ( f_{i}\epsilon_i )
  \right\|_\infty
  \|\widehat \beta - \beta_{0}\|_{1}   +   |\Gn \{ a_{i} \epsilon_i\} | \to_P 0
 \end{eqnarray*}
by condition (\ref{RateConditionI}) since $|\Gn \{ a_{i} \epsilon_i\} |\lesssim_P c_s\sigma_\epsilon$ and $\sigma_\epsilon$ is bounded above uniformly in $n$.

Step 4. This step establishes
consistency of the variance estimator in the homoscedastic case.
Since $\sigma^2_\epsilon$ and  $Q_n=\En[A_i A_i']$ are bounded away from zero
and from above uniformly in $n$, it suffices to show $\hat \sigma^2_\epsilon - \sigma^2_\epsilon \to_P 0$
and  $\En [\widehat A_i\widehat A_i'] - \En[A_i A_i'] \to_P 0$.

Indeed,
$\hat\sigma^2_\epsilon = \En[(\epsilon_i-d_i'(\widehat\alpha^*-\alpha_0))^2] =
\En[\epsilon_i^2]+2\En[\epsilon_id_i'(\alpha_0-\widehat\alpha^*)]
+\En[(d_i'(\alpha_0-\widehat\alpha^*))^2]$ so that
$\En[\epsilon_i^2]- \sigma^2_\epsilon \to_P 0$ by Chebyshev inequality since $\Ep[\epsilon_i^4]$ is bounded
uniformly in $n$, and the remaining
terms converge to zero in probability since $\widehat \alpha^* - \alpha_0 \to_P
0$, $\|\En[d_i\epsilon_i]\| \lesssim_P 1$ by Step 0.

Next, note that
$$ \|\En [\widehat A_i\widehat A_i'] - \En [A_i A_i'] \| = \|\En
[A_i (\widehat A_i-A_i)' + (\widehat A_i-A_i)A_i'] + \En
[(\widehat A_i-A_i) (\hat A_i-A_i)'] \|
$$
which is bounded up to a constant by
$$
\|\widehat A_{i}-A_{i}\|_{2,n} \|A_i\|_{2,n} + \|\widehat A_{i}-A_{i}\|^2_{2,n} \to_P 0
$$
since $\|\widehat A_{i}-A_{i}\|^2_{2,n} = \|\widehat D_{i}-D_{i}\|^2_{2,n}=o_P(1)$ by  (\ref{RateConditionI}), and  $\|A_i\|_{2,n} \lesssim 1$ holding by Step 0.\qed

\section{Proof of Theorem 2}

Step 0.  The step here is identical to Step 0 of the proof of Theorem 1, whereby
we introduce additional indices $a$ and $b$ on all variables; and $n$ gets replaced
by either $n_a$ or $n_b$.

Step 1. We have that by $\Ep[\epsilon^k_i|A^k_i]=0$ for both $k=a$ and $k= b$,
\begin{eqnarray*}
\sqrt{n_k}(\widehat \alpha_k - \alpha_0) &= & \Enk [\widehat A^k_i d_i^k{}']^{-1} \sqrt{n} \En_j [\widehat A^k_i \epsilon^k_i]  \\
& = & \{ \Enk [\widehat A^k_i d^k_i{}']\}^{-1} {\Gnk} [\widehat A^k_i \epsilon^k_i]  \\
& = & \{ \Enk [A^k_i A^k_i{}'] + o_P(1)\}^{-1} \left( {\Gnk} [A_i^{k} \epsilon_i^k{}] + o_P(1) \right)
 \end{eqnarray*}
where
\begin{eqnarray}
 \Enk [\widehat A^k_id^k_i{}'] = \Enk[A^k_iA^k_i{}'] + o_P(1) \label{eq: to show a} \\
{ \Gnk} [\widehat A^k_i \epsilon^k_i] = {\Gnk} [A^k_i \epsilon^k_i] + o_P(1)\label{eq: to show b}.
 \end{eqnarray}
where  (\ref{eq: to show a}) follows similarly to Step 2 in the proof of Theorem 1 and condition (\ref{RateConditionSplitIV}).
The relation (\ref{eq: to show b}) follows from Chebyshev inequality and
$$
E[ \| {\Gnk} [(\widehat A^k_i - A^k_i) \epsilon^k_i] \|^2| k^c]  \lesssim  \sigma^2_{\epsilon} \|(\widehat A^k_i - A^k_i)  \|_{2,n_k}^2 \to_P 0$$
where we used that $(\widehat A^k_i - A^k_i), 1 \leq i \leq n_k$ by construction are independent of $\epsilon^k_i, 1 \leq i \leq n_k $ and that  $\|(\widehat A^k_i - A^k_i)  \|_{2,n_k} \leq \|(\widehat D^k_i - D^k_i)  \|_{2,n_k} \to_P 0$,
where $E[\cdot |k^c]$ denotes the estimate computed conditional on the sample $k^c$, where $k^c= \{a,b\}\setminus k$.

By assumption eigenvalues of $\Enk [A^k_i A^k_i{}']$ and $\sigma_\epsilon$ are bounded away and above from zero, and so we can conclude that
\begin{eqnarray*}
\sqrt{n_k}(\widehat \alpha_k - \alpha_0)
& = & \{ \Enk [A^k_i A^k_i{}']\}^{-1} \left( {\Gnk} [A_i^{k} \epsilon_i^k{}] + o_P(1) \right) \\
& = & \{ \Enk [A^k_i A^k_i{}']\}^{-1/2} \sigma_{\epsilon} Z_k +  o_P(1)
 \end{eqnarray*}
where $Z_a$ and $Z_b$ are two independent $N(0,I)$ vectors;  and also note that $\sqrt{n_k}(\widehat \alpha_k - \alpha_0) = O_P(1)$, for $k=a,b$.

Step 3.  Now putting together terms we get
\begin{eqnarray*}
\sqrt{n}(\widehat \alpha_{ab} - \alpha_0) & = &
( (n_a/n)  \Ena [\hat A_i^a  \hat A_i^{a}{}']  +  (n_b/n) \Enb [\hat A_i^b  \hat A_i^{b}{}']   )^{-1} \times \\
& \times &  ( (n_a/n)  \Ena [\hat A_i^a  \hat A_i^{a}{}']  \sqrt{n} (\widehat \alpha_a - \alpha_0) +   (n_b/n) \Enb  [\hat A_i^b  \hat A_i^{b}{}'] \sqrt{n} (\widehat \alpha_b - \alpha_0)   ) \\
& = &
( (n_a/n)  \Ena [ A_i^a   A_i^{a}{}']  +  (n_b/n) \Enb [ A_i^b   A_i^{b}{}']   )^{-1} \times \\
& \times &  ( (n_a/n)  \Ena [ A_i^a  A_i^{a}{}']  \sqrt{n} (\widehat \alpha_a - \alpha_0) +   (n_b/n) \Enb  [ A_i^b   A_i^{b}{}'] \sqrt{n} (\widehat \alpha_b - \alpha_0)   ) + o_P(1) \\
& = & \{ \En [A_i A_i{}']\}^{-1} \times \\
& \times  &  ( \frac{1}{2} \Ena [ A_i^a   A_i^{a}{}']^{1/2}  \sqrt{2}\sigma_{\epsilon} Z_a
+  \frac{1}{2} \Enb[ A_i^b   A_i^{b}{}']^{1/2} \sqrt{2} \sigma_{\epsilon} Z_b  )+ o_P(1) \\
& = &  \{ \En [A_i A_i{}']\}^{-1}  N(0, \frac{\sigma^2_\epsilon}{2} \Ena[ A_i^a   A_i^{a}{}'] +  \frac{\sigma^2_\epsilon}{2} \En_{b}[ A_i^b   A_i^{b}{}']   )  + o_P(1) \\
& = &  \{ \En [A_i A_i{}']\}^{-1}  N(0,  \sigma^2_\epsilon  \En [ A_i   A_i'])  + o_P(1) \\
& = &    N(0,   \sigma^2_\epsilon \{ \En [ A_i   A_i']\}^{-1})  + o_P(1)
 \end{eqnarray*}
The conclusions now follow as in the proof of Theorem 1.

Step 3.  This step is similar to Step 4 in the proof of Theorem 1.\qed

\section{Proof of Lemma 2 (Rates for LASSO and Post-LASSO)}

Note that $\| \widehat D_i - D_i \|_{2,n} \leq\|f_i'(\widehat \beta - \beta_0)\|_{2,n} + \|a(x_i)\|_{2,n} \leq \|f_i'(\widehat \beta - \beta_0)\|_{2,n} + c_s$. Let $\delta :=\widehat \beta - \beta_0$ and $c_0 = (c+1)/(c-1)$.

First consider the LASSO estimator.
By optimality of the LASSO estimator and expanding $\widehat Q(\widehat \beta)-\widehat Q(\beta_0)$, if $ \lambda \geq c n \|S\|_\infty$
we have
\begin{equation}\label{Eq:Basic} \begin{array}{rcl}\|f_i'\delta\|_{2,n}^2  & \leq & \displaystyle \frac{\lambda}{n}\(\|\delta_T\|_1 - \|\delta_{T^c}\|_1\) + \|S\|_{\infty}\|\delta\|_1 + 2c_s\|f_i'\delta\|_{2,n}\\
& \leq & \displaystyle \(1 + \frac{1}{c}\) \frac{\lambda}{n} \|\delta_T\|_1 - \(1-\frac{1}{c}\)\frac{\lambda}{n}\|\delta_{T^c}\|_1 +2c_s\|f_i'\delta\|_{2,n}.\\
\end{array}
\end{equation}
That yields that either $\|f_i'\delta\|_{2,n} \leq 2c_s$ or that
$\|\delta_{T^c}\|_1 \leq c_0\|\delta_T\|_1$. As shown Lemma 1 of \cite{BC-PostLASSO} (which is based on \cite{BickelRitovTsybakov2009}) if $ \lambda \geq c n \|S\|_\infty$ we have
$$ \|f_i'\delta\|_{2,n} \leq \frac{2\lambda\sqrt{s}}{n \kappa_{c_0}} + 2c_s \lesssim \sigma_v\sqrt{\frac{s\log (p/\conflvl)}{n}}$$ since $\kappa_{c_0}$ is bounded away from zero by condition RE as $n\to \infty$ and the choice of penalty level (\ref{Def:LambdaLASSO}). Note that under the penalty choice (\ref{Def:LambdaLASSO}) we have that $ \lambda \geq c n \|S\|_\infty$ with probability $1-\conflvl \to 1$ since $\gamma = o(1)$.

Under our conditions, we can invoke sparsity bound for LASSO by Theorem 5 in \cite{BC-PostLASSO}. We have that
$$ \| \delta\|_0 \lesssim_P s.$$
Therefore, we have $$\|\delta\|_2 \leq \|f_i'\delta\|_{2,n} / \sqrt{\semin{\| \widehat \delta\|_0}} \lesssim_P \|f_i'\delta\|_{2,n}$$ since for any fixed $C>0$, $\semin{Cs}$ is bounded away from zero as $n$ grows by condition SE.

To establish the last result, if $\|\delta\|_0 \lesssim_P s$, it follows that $\|\delta\|_1 \leq \sqrt{\|\delta\|_0} \|\delta\|_2\lesssim_P \sqrt{s}
\|\delta\|_2$.

\comment{Alternatively, consider two cases. First, assume $\|
\delta_{T^c}\|_1 \leq 2c_0 \| \delta_T\|_1.$ In this
case, by definition of the restricted eigenvalue, we have
$$ \|\delta\|_1 \leq (1+2c_0) \|\delta_T\|_1 \leq (1+2c_0)\sqrt{s}\|f_t'\delta\|_{2,n}/\kappa_{2c_0} $$
and the result follows by applying the first bound to
$\|f_t'\delta\|_{2,n}$  since $c_0 > 1$.

On the other hand, consider the case that $\|\delta_{T^c}\|_1 > 2c_0 \|\delta_T\|_1$ which would already imply $\|f_t'\delta\|_{2,n} \leq 2c_s$. Moreover, the relation (\ref{Eq:Basic}) implies that
$$\begin{array}{rcl}
 \|\delta_{T^c}\|_1 & \leq & c_0 \|\delta_T\|_1 + \frac{c}{c-1} \frac{n}{\lambda}\|f_t'\delta\|_{2,n}(2c_s - \|f_t'\delta\|_{2,n})\\
 & \leq & c_0 \|\delta_T\|_1 + \frac{c}{c-1}\frac{n}{\lambda} c_s^2\\
& \leq & \frac{1}{2} \| \delta_{T^c}\|_1 + \frac{c}{c-1}\frac{n}{\lambda} c_s^2\\
\end{array}
$$
 Thus,
$$\| \delta\|_1 \leq \(1 + \frac{1}{2c_0}\) \| \delta_{T^c}\|_1 \leq \(1 + \frac{1}{2c_0}\)\frac{2c}{c-1}\frac{n}{\lambda} c_s^2, $$
and the result follows since $nc_s^2/\lambda \lesssim_P s/\sqrt{n}$ since $c_s \lesssim_P \sqrt{s/n}$ and $\sqrt{n} \lesssim \lambda$ because $\conflvl = o(1)$.
}

The proof for the post-LASSO estimator follows from Theorem 6 in \cite{BC-PostLASSO} and the sparsity bound for LASSO in Theorem 5  in \cite{BC-PostLASSO}.

\section{Proof of Lemma 3 (Rates for $\LASSO$ and Post-$\LASSO$)}

The proof for the $\LASSO$ and the Post-$\LASSO$ estimator follows from \cite{BCW-SqLASSO,BCW-SqLASSO2}.

\section{Proof of Theorem 3}

First note that  by a union bound and tail properties of the Gaussian random variables, see e.g. \cite{BickelRitovTsybakov2009},
$$ \left \|\Gn ( f_{i} \epsilon_i  ) \right \|_{\infty} \lesssim_P \sigma_\epsilon\sqrt{\log p} $$ since $\epsilon_i\sim N(0,\sigma_\epsilon^2)$ and $\En[f_{ij}^2] = 1$ for $j=1,\ldots,p$.
Under the condition that $s^2\log^2 p = o(n)$, the result follows by applying the rates in Lemma 1 (for LASSO and Post-LASSO) and Lemma 2 (for $\LASSO$ and Post-$\LASSO$) to verify condition (\ref{RateConditionI}) in Theorem \ref{lemma: lead lemma}.

\section{Proof of Theorem 4}

For every observation $i$ in the subsample $k$ we have
$$ D_i = f_i'\beta_0 + a(x_i) = f_i^k{}'H_k\beta_0+a(x_i), \ \ \|H_k\beta_0\|_0 \leq s$$
so that $H_k\beta_0$ is the target vector for the renormalized subsample $k$.

Under our conditions, we can invoke sparsity bound for LASSO by Theorem 5 in \cite{BC-PostLASSO} and for $\LASSO$ by \cite{BCW-SqLASSO2}. In either case, we have that for $\delta = \widehat \beta^k - H_k\beta_0$, $k=a,b$,
$$ \| \delta\|_0 \lesssim_P s.$$
Therefore, by condition SE, we have for $M=\Ena[f_i^af_i^a{}']$, $\Enb[f_i^bf_i^b{}']$ or $\En[f_if_i{}']$, that with probability going to $1$, for $n$ large enough we have $$0 < \kappa' \leq \semin{\|\delta\|_0} \leq \semax{\|\delta\|_0} \leq \kappa''<\infty.$$

Therefore, we have $\|H_{k^c}\beta_0 - \widehat \beta^{k^c}\|_0\lesssim_P s$ and
\begin{equation}\label{ExtraStep} \begin{array}{rl}
\|D^k_i - \widehat D_i^k\|_{2,n_k} & = \|f_i^k{}'H_k\beta_0 + a(x_i^k) - f_i^k{}'H_kH_{k^c}^{-1} \widehat \beta^{k^c}\|_{2,n_k} \\
& = \|f_i^k{}'H_kH_{k^c}^{-1}(H_{k^c}\beta_0 - \widehat \beta^{k^c})\|_{2,n_k} + \|a(x_i^k)\|_{2,n_k} \\
& \leq \sqrt{ \kappa''/\kappa' } \|H_kH_{k^c}^{-1}\|_\infty \|f_i^{k^c}{}'(\widehat \beta^{k^c}- H_{k^c}\beta_0^{k^c})\|_{2,n_{k^c}} + c_s\sqrt{n/n_k}\\
\end{array}\end{equation}
where the last inequality holds with probability going to 1. Moreover, note that $\|H_kH_{k^c}^{-1}\|_\infty \leq \sqrt{\semax{1}/\semin{1}} \lesssim 1$ by condition SE.

Then, under (\ref{eq: sparse instrument}) and $s\log p = o(1)$, the result is an immediate consequence of Theorem 2 since (\ref{RateConditionSplitIV}) holds by (\ref{ExtraStep}) combined with Lemma 1 (for LASSO and Post-LASSO) and Lemma 2 (for $\LASSO$ and Post-$\LASSO$) that imply
$$ \|f_i^k{}'(\widehat \beta^{k}- H_k\beta_0)\|_{2,n_{k}} = o_P(1), \ \ \ k=a,b.$$

\bibliographystyle{plain}
\bibliography{mybib}

\pagebreak

\begin{figure}
    \includegraphics[width=5.8in]{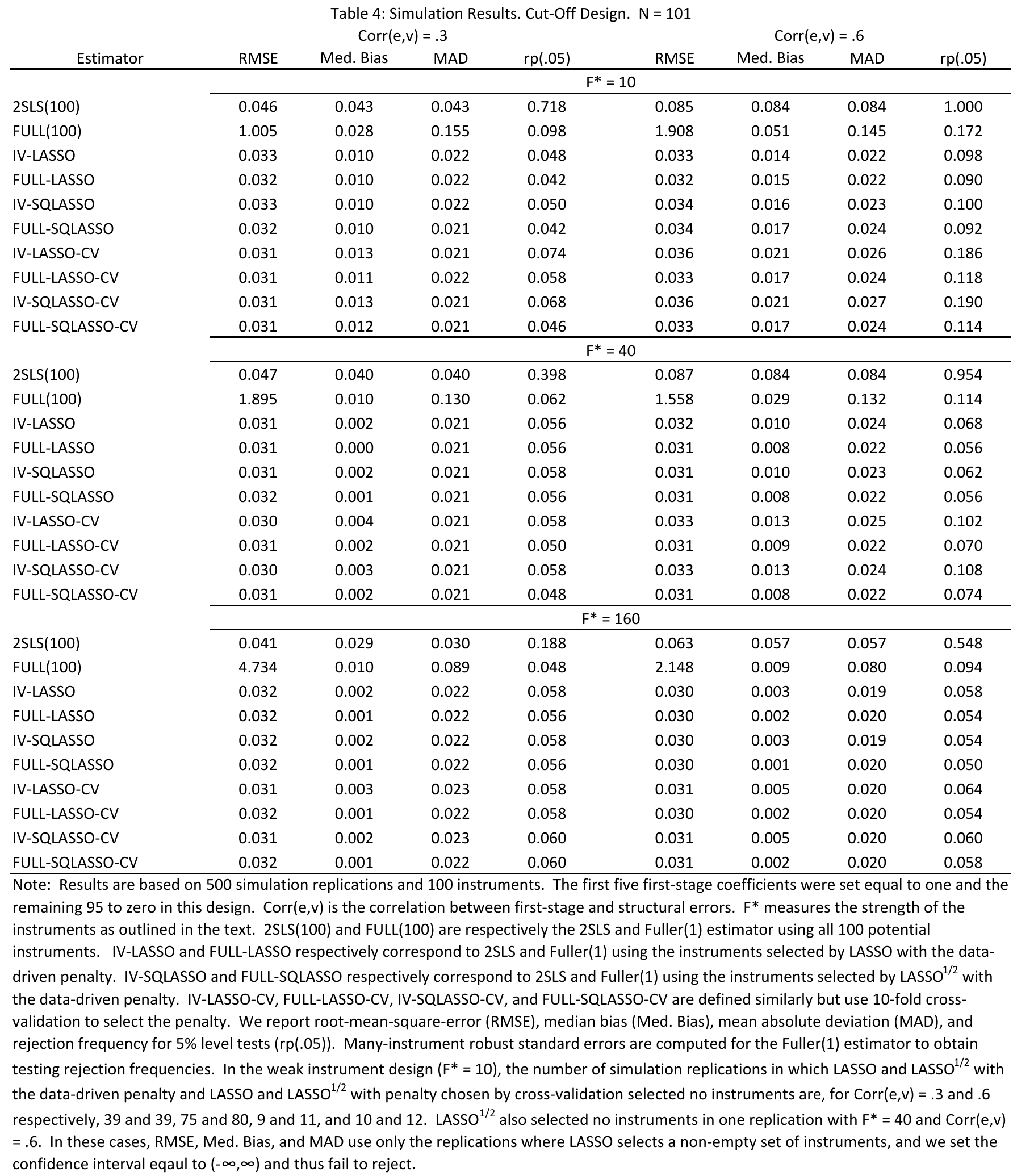}
    \label{fig:table1}
\end{figure}

\pagebreak

\begin{figure}
    \includegraphics[width=5.8in]{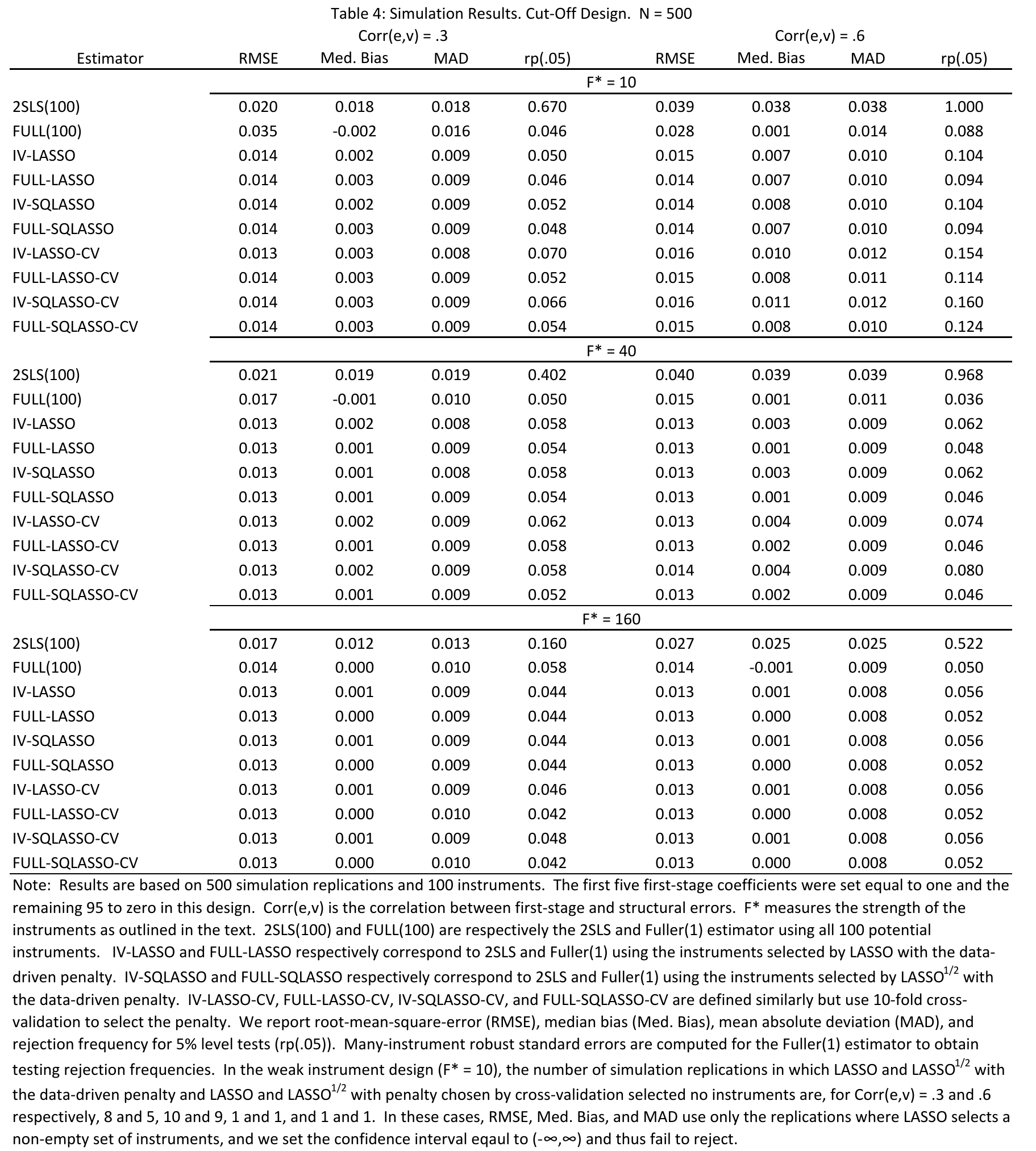}
    \label{fig:table2}
\end{figure}

\pagebreak

\begin{figure}
    \includegraphics[width=5.8in]{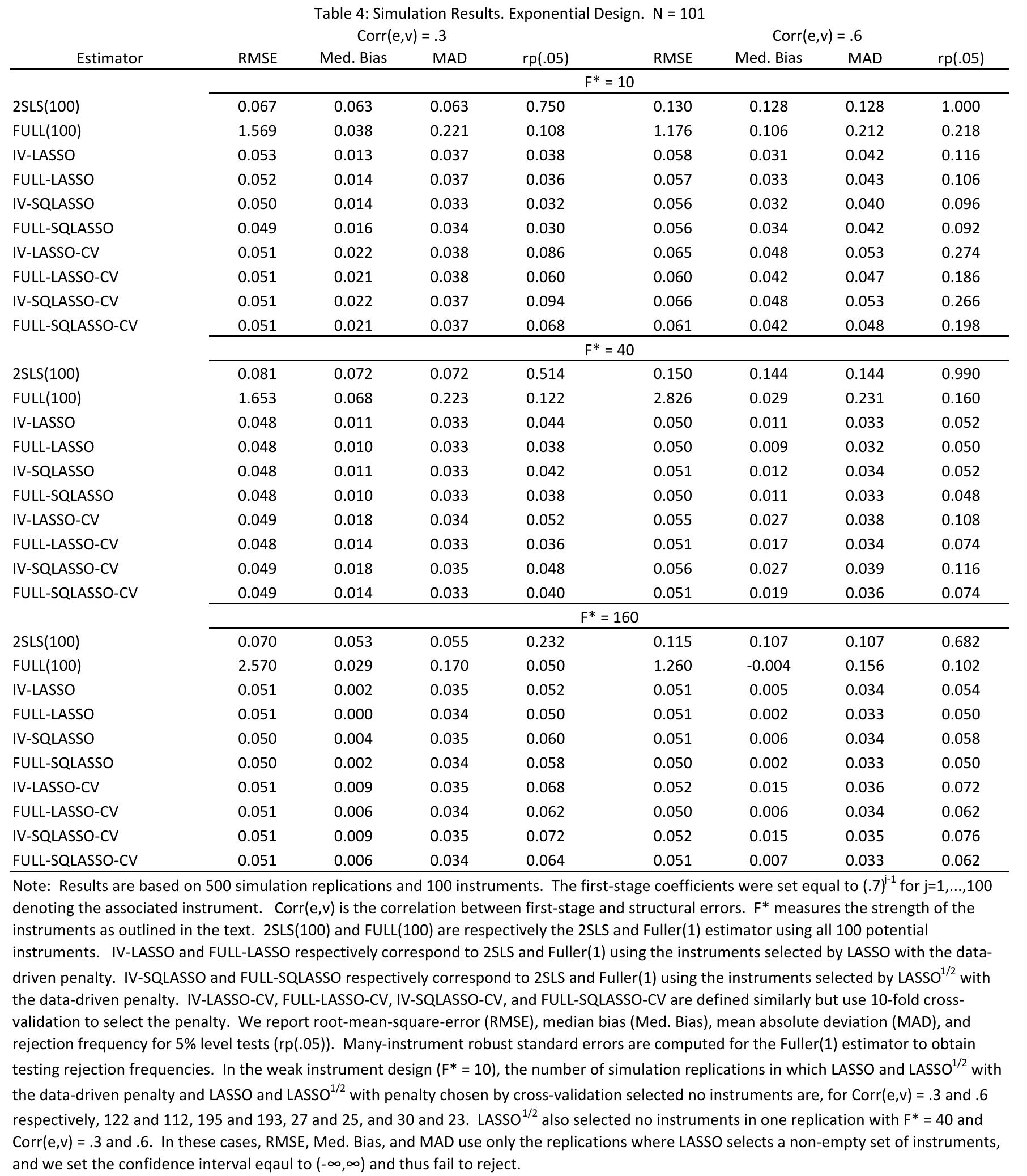}
    \label{fig:table4}
\end{figure}

\pagebreak

\begin{figure}
    \includegraphics[width=5.8in]{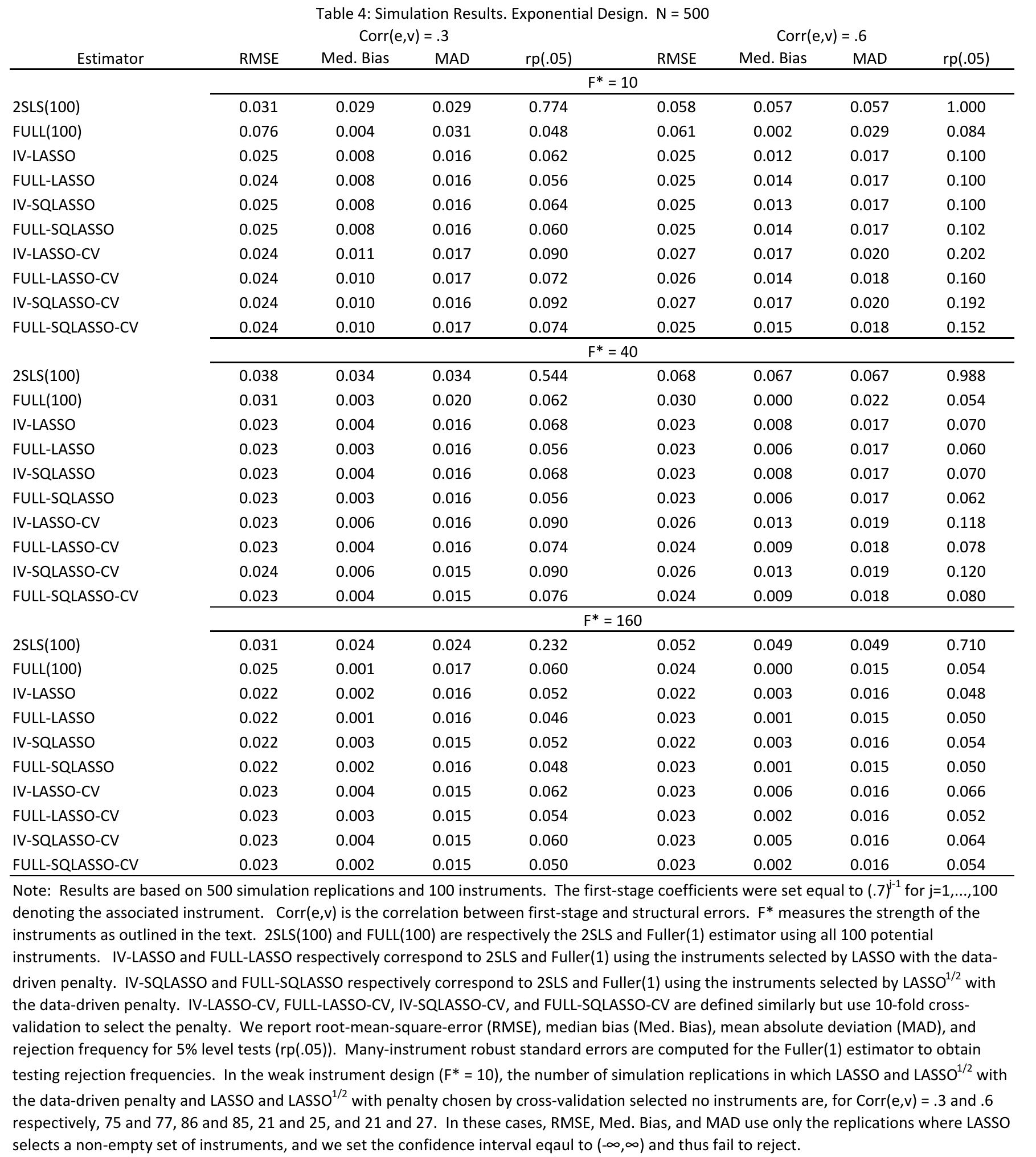}
    \label{fig:table5}
\end{figure}

\begin{figure}
    \includegraphics[width=5.8in]{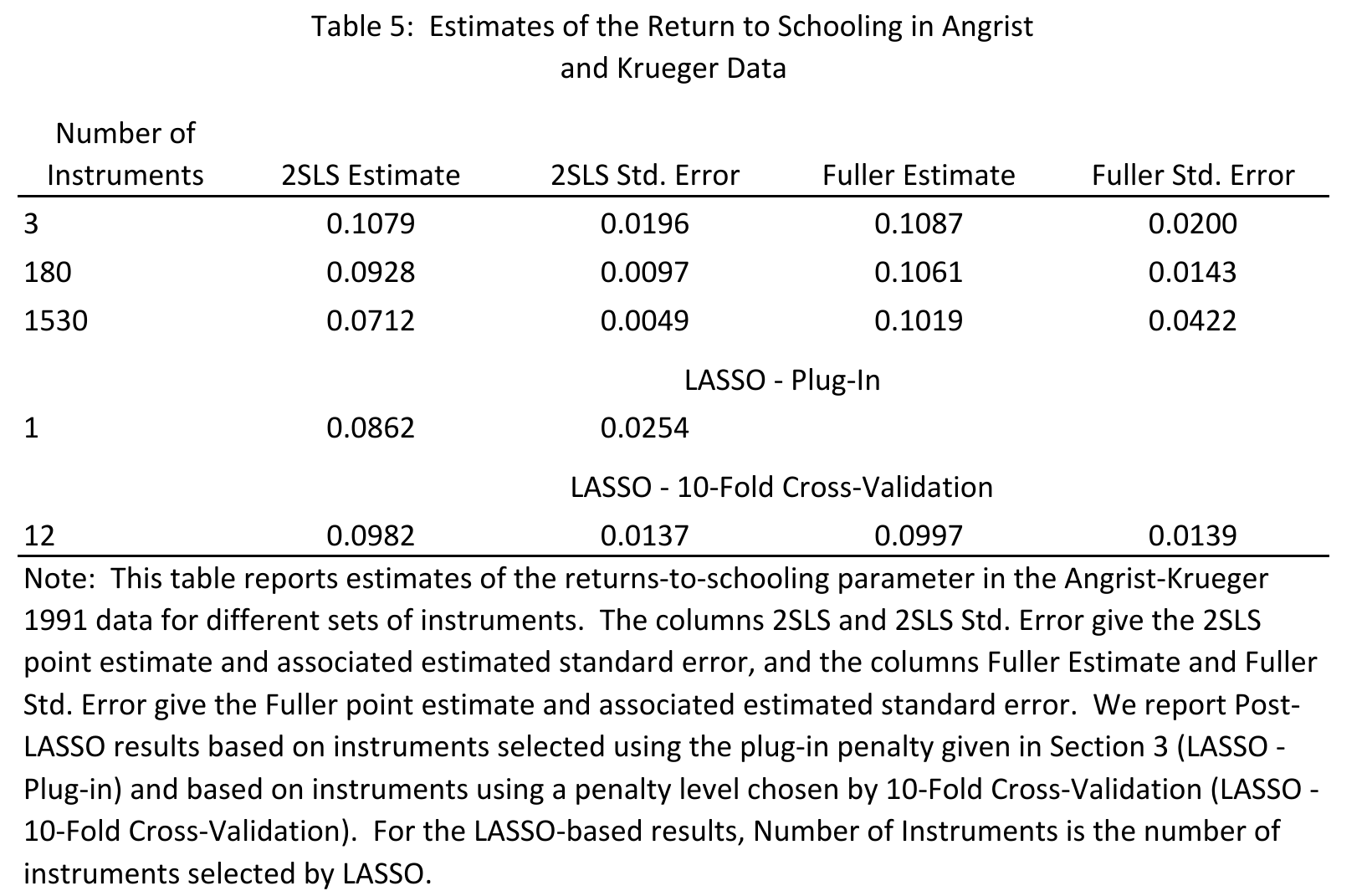}
    \label{fig:table6}
\end{figure}

\end{document}